\documentclass[runningheads]{llncs}
\usepackage[T1]{fontenc}
\usepackage{booktabs}
\usepackage{graphicx}
\usepackage{amsmath}
\usepackage{subfloat}
\usepackage{makecell}
\usepackage{color}
\usepackage{float}
\usepackage{newtxmath}
\usepackage{setspace}
\usepackage{bm}
\usepackage{graphicx}  
\usepackage{float}  
\usepackage{subfigure}  
\usepackage{hyperref}

\def\onedot{$\mathsurround0pt\ldotp$}
 
\def\ie{\emph{i.e}\onedot}

\begin{document}
\title{GENERATING HIGH-QUALITY SYMBOLIC MUSIC 
USING FINE-GRAINED DISCRIMINATORS}

\author{Zhedong Zhang\inst{1} \and
Liang Li\inst{2} \and
Jiehua Zhang\inst{3} \and
Zhenghui Hu\inst{4} \and
Hongkui Wang\inst{1} \and
Chenggang Yan\inst{1} \and
Jian Yang\inst{5} \and
Yuankai Qi\inst{5} 
}
\authorrunning{Z. Zhang et al.}
%
\institute{
Hangzhou Dianzi University, Hangzhou, Zhejiang, China \and
Institute of Computing Technology, Chinese Academy of Sciences, Beijing, China \and
Xi'an Jiaotong University, Xi'an, Shan Xi, China \and
Hangzhou Innovation Institute, Beihang University, Hangzhou, Zhejiang, China \and
Macquarie University, Sydney, New South Wales, Australia
}
%
\maketitle              
%
\begin{abstract}
Existing symbolic music generation methods usually utilize discriminator to improve the quality of generated music via global perception of music.
However, considering the complexity of information in music, such as rhythm and melody, a single discriminator cannot fully reflect the differences in these two primary dimensions of music.
In this work, we 
propose to decouple the melody and rhythm from music, and design corresponding fine-grained discriminators to tackle the aforementioned issues.
%
%
Specifically, equipped with a pitch augmentation strategy, the melody discriminator discerns the melody variations presented by the generated samples.
By contrast, the rhythm discriminator, enhanced with bar-level relative positional encoding, focuses on 
the 
velocity of generated notes.
Such a design allows the generator to be more explicitly aware of which aspects should be adjusted in the generated music, making it easier to mimic human-composed music.
Experimental results on the POP909 benchmark demonstrate the favorable performance of the proposed method compared to several state-of-the-art methods
in terms of both objective and subjective metrics.
More demos are available at \hyperref[https://zzdoog.github.io/fg-discriminators/]{https://zzdoog.github.io/fg-discriminators/}.
\end{abstract}
\section{Introduction}
\label{intro}
Due to the high-level representation of music based on Musical Instrument Digital Interface (MIDI) and its variants, symbolic music generation models do not need to learn how to create the sounds of various instruments so that they can focus more on the music itself~\cite{survey,wavenet,jukebox}.
Since the high-level discrete tokens of music are similar to words of text, transformer-based models~\cite{music_transformer,CP,pop_music,popmag} have been widely applied in symbolic music generation, and towards the goal of generating high-quality music in recent years.
Most symbolic music generation models are trained to maximize the likelihood of observed sequences. 
These methods can learn the patterns of discrete token sequences and ensure statistical consistency, but they may suffer from noticeable quality degradation when generating complex music sequences due to exposure bias~\cite{transformer-gan}.

 
\begin{figure}
\centering
\includegraphics[width=0.8\linewidth]{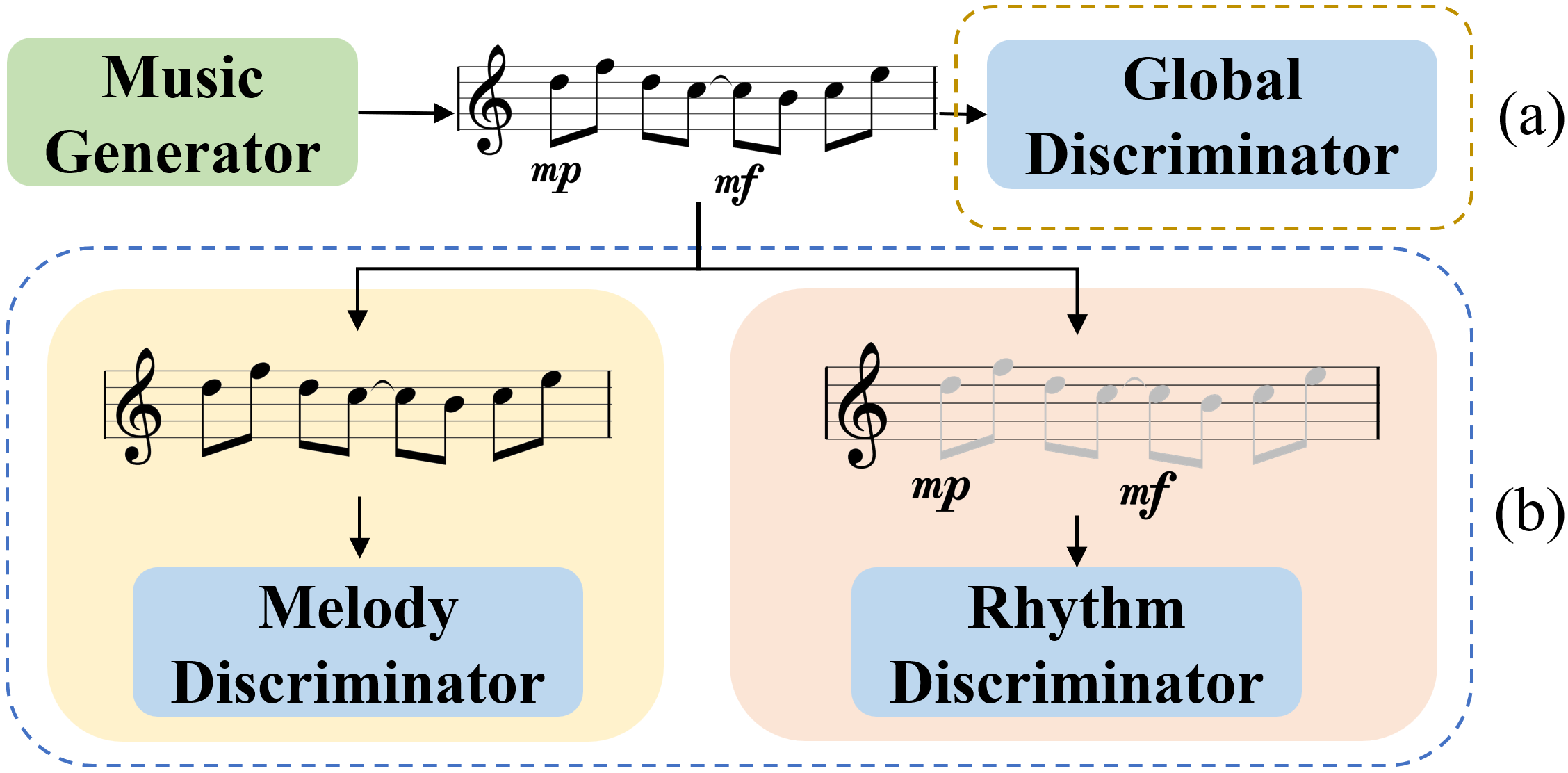}
\caption{(a) Main structure of conventional GAN-based method with coarse-grained global discriminator. (b) The structure of proposed fine-grained discriminators architecture.
} 
\label{intro_fig}
\end{figure}

Some studies~\cite{transformer-gan,adtransformer,musegan} have attempted to address the aforementioned issues by introducing adversarial loss~\cite{gan}. 
They enhance the generative model by utilizing the feedback from the discriminator based on the discriminator's discernment on generated and real music.
%
Despite the progress, 
their discriminators cannot explicitly reflect the defects in terms of two important music properties: melody and rhythm, due to the lack of corresponding designs.
According to music perception theory, melody and rhythm are two primary dimensions of music~\cite{melodyandrhythm}. They respectively represent the arrangement of musical pitches in a particular order and the progression patterns of notes, which constitute the core of music composition~\cite{performance}.
Well-sounding music should feature a stable melody with rich variation, supported by a rhythmic framework that maintains smooth and varied progressions~\cite{melodyandrhythm}.
Lacking an effective targeted model, the quality of music generated by existing methods is limited.
%

To address the above problems, we propose a novel architecture with fine-grained discriminators for symbolic music generation, as shown in Fig.~\ref{intro_fig}.
Aiming to provide fine-grained adversarial feedback to the generator, we first design a decoupling module to well disentangle the melody and rhythm information from music.
Specifically, we mask all the note velocity and note pitch tokens with the same token \texttt{[Mask]} in the sequence respectfully to extract melody and rhythm information from the original music sequence.
After decoupling, we design the corresponding fine-grained melody and rhythm discriminator for the generator.
To discriminate whether the melodies of generated music closely resemble real data, 
a pitch augmentation strategy is used in the melody discriminator to reduce the impact of the absolute pitch.
Correspondingly, we design a fine-grained rhythm discriminator elaborately.
By devising a bar-level relative positional encoding to enhance the discriminator to better capture the rhythm pattern within the local structure.

The contributions of this paper are summarized below:
\begin{itemize}
\item[$\bullet$] We propose a fine-grained discriminators architecture for melody and rhythm respectfully in symbolic music generation domain, which is more aligned with music perception theories. 
\item[$\bullet$] 
We design a melody-rhythm decoupling module for symbolic music and incorporate pitch enhancement strategies and bar-level relative position encoding to enhance the corresponding fine-grained discriminators, providing elaborate feedback to the generator.
\item[$\bullet$] Extensive experiments show the favorable performance of our method in terms of both objective metrics and subjective listening tests. More generated examples are available at supplementary materials.
\end{itemize}


\begin{figure*}[!h]
\setlength{\abovecaptionskip}{0.cm}
\includegraphics[width=\linewidth]{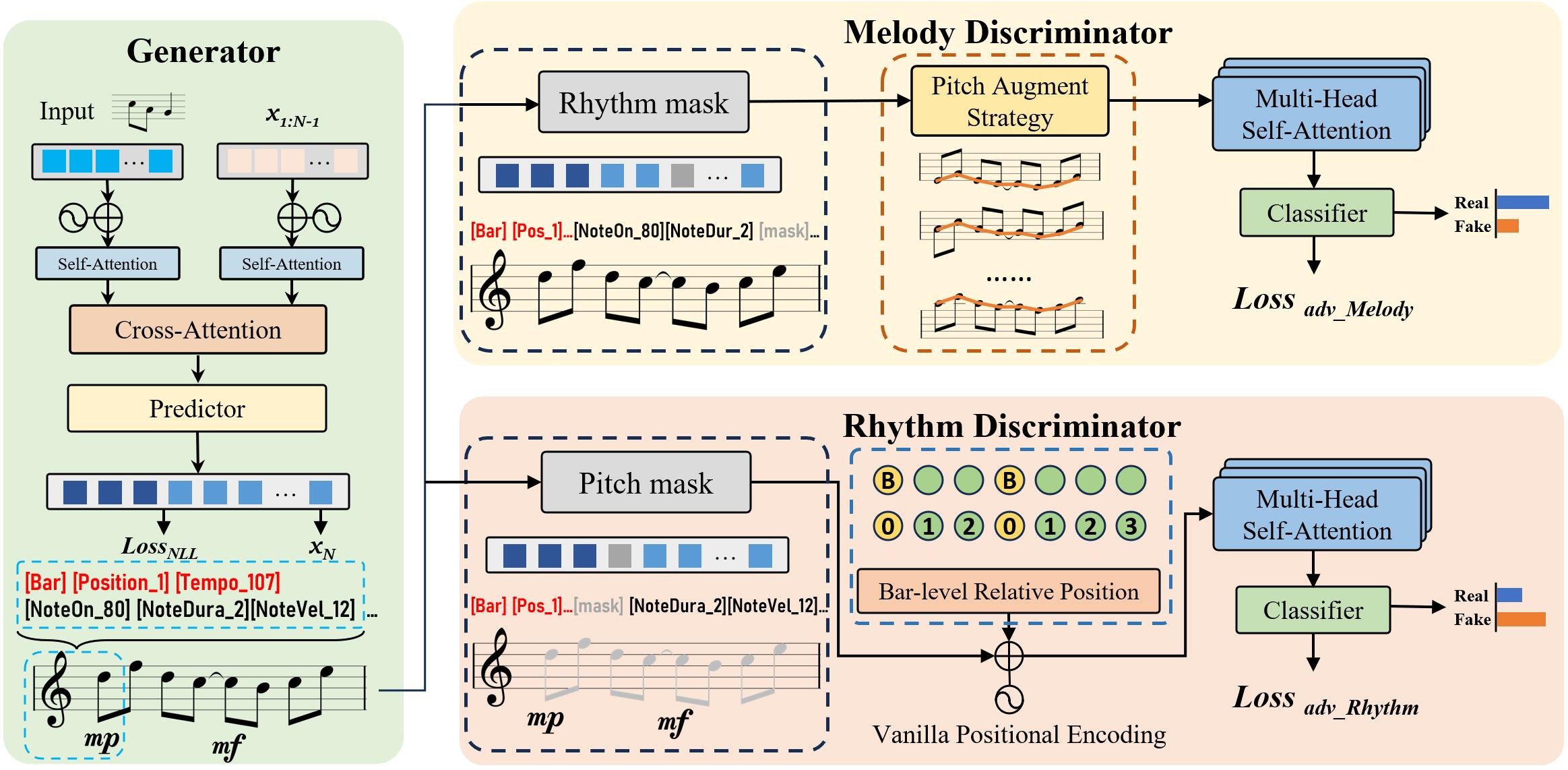}
\caption{
Main framework of the proposed symbolic music generation model, consists of three main components: a music generator and two fine-grained discriminators --- rhythm discriminator and melody discriminator.} \label{method}
\end{figure*}
\section{METHODOLOGY}
\label{sec:format}
The proposed model consists of an auto-regressive symbolic music generator and two fine-grained discriminators as shown in Fig.~\ref{method}.
First, the generator takes a representative condition music sequence $c$ as the input, and attempts to generate whole music sequence align with the condition.
%
Then, during the generator optimization, the fine-grained melody and rhythm discriminators provide more precise feedback to the generator by decoupling and analyzing the output of the generator.
Simultaneously, the fine-grained discriminators continually enhance their discriminatory abilities relying on samples generated by the evolving generator to provide further feedback to the generator.
%
The value function of the generator and fine-grained melody and rhythm discriminators are defined as follows:
\begin{equation}\label{target function}
    \setlength{\abovedisplayskip}{9pt}
    \setlength{\belowdisplayskip}{9pt}
    \begin{split}
    &\min_{G_{\theta}}\max_{D_{m},D_{r}}V=\{\mathbb{E}[logD_{m}(s_{r})]+\mathbb{E}[logD_{r}(s_{r})]+ \\
    &\mathbb{E}[log(1-D_{m}(G_{\theta}(c)))]+\mathbb{E}[log(1-D_{r}(G_{\theta}(c)))]\},
    \end{split}
\end{equation}
where the $G$, $D_{m}$, and $D_{r}$ denote the generator,melody discriminator,
and rhythm discriminator respectively.
$\theta$ and $s_{r}$ denote the parameter of the generator and real sample from the dataset respectively.

\vspace{-0.3cm}
\subsection{Generator}
\label{Generator}
\vspace{-0.1cm}
We adopt the seq2seq symbolic music generation transformer model~\cite{theme} as our generator.
It takes condition music sequence as input and generates a complete and harmonious music composition that aligns with the input.
The condition music sequence is the thematic material of each music composition, implies the main idea of the whole composition, retrieved from the complete music by clustering algorithm~\cite{theme}.
The overall loss function of the generator as follows: 
\begin{equation}
\setlength{\abovedisplayskip}{6pt}
\setlength{\belowdisplayskip}{6pt}
\mathcal{L}_{G} = \mathcal{L}_{NLL}+\alpha\cdot\mathcal{L}_{adv\_{Melody}}+\beta\cdot\mathcal{L}_{adv\_{Rhythm}},
 \label{total loss of the model}
\end{equation}
\begin{equation}
\setlength{\abovedisplayskip}{6pt}
\setlength{\belowdisplayskip}{6pt}
\mathcal{L}_{NLL} = \sum_{n=1}^N-logP(x_n|\theta,x_{1:n-1}, c),
\end{equation}
where the $\alpha$ and $\beta$ are pre-defined hyper-parameters.
Details of the two adversarial losses are in the following sections.
Note that our fine-grained discriminators architecture applies equally to other state-of-the-art music generation methods.

\subsection{Melody Discriminator with Pitch Augmentation Strategy}
\label{ssec:subhead}

Melody is one of the primary properties of music. 
It provides a tuneful and recognizable musical line that serves as a focal point for listeners. 
The arrangement of pitches in a particular order and duration forms the melody~\cite{pitchtheory}.
Traditional NLL-trained models perform poorly in generating long and harmonious melodies due to the lack of specific guidance.
To deal with this issue, we propose a melody discriminator with a pitch augmentation strategy to facilitate the discrimination of the melody in generated music.

%
First, we decouple the melody information from symbolic music
by replacing all the \texttt{[Note-Velocity]} tokens with the \texttt{[mask]} token.
Then, to enhance our melody discriminator, we augment the original data via uniformly raising or decreasing the absolute pitch of all original notes to simulate the melody in different voice parts, as shown in Fig.~\ref{method} top right.
%
%
All these decoupled melody data are fed into the melody discriminator which uses an encoder-only transformer with a multi-head self-attention mechanism as backbone~\cite{transformer}.
%
During the adversarial training process, the adversarial loss from the melody discriminator and back-propagation gradients to the melody discriminator are formulated as :
\begin{equation} \label{pitch discriminator adversarial loss}
\setlength{\abovedisplayskip}{6pt}
\setlength{\belowdisplayskip}{6pt}
\begin{split}
\mathcal{L}_{adv\_{Melody}}=\frac{1}{N}\sum_{i=1}^{N}[log(1-D_{m}(G_{\theta}(c^{(i)})))], \\
\nabla_{\theta_{m}}\frac{1}{N}\sum_{i=1}^{N}[log(D(s_r^{(i)})+log(1-D_{m}(G_{\theta}(c^{(i)})))],
\end{split}
\end{equation}
where $\theta_{m}$ denotes the parameter
of melody discriminator, $s_{r}^{(i)}$ and $c_{r}^{(i)}$ denote as $i$-th ground truth and conditional input.

\subsection{Rhythm Discriminator with Bar-level Relative Positional Encoding}
\label{ssec:subhead}

\begin{figure}
\centering
\includegraphics[width=0.8\linewidth]{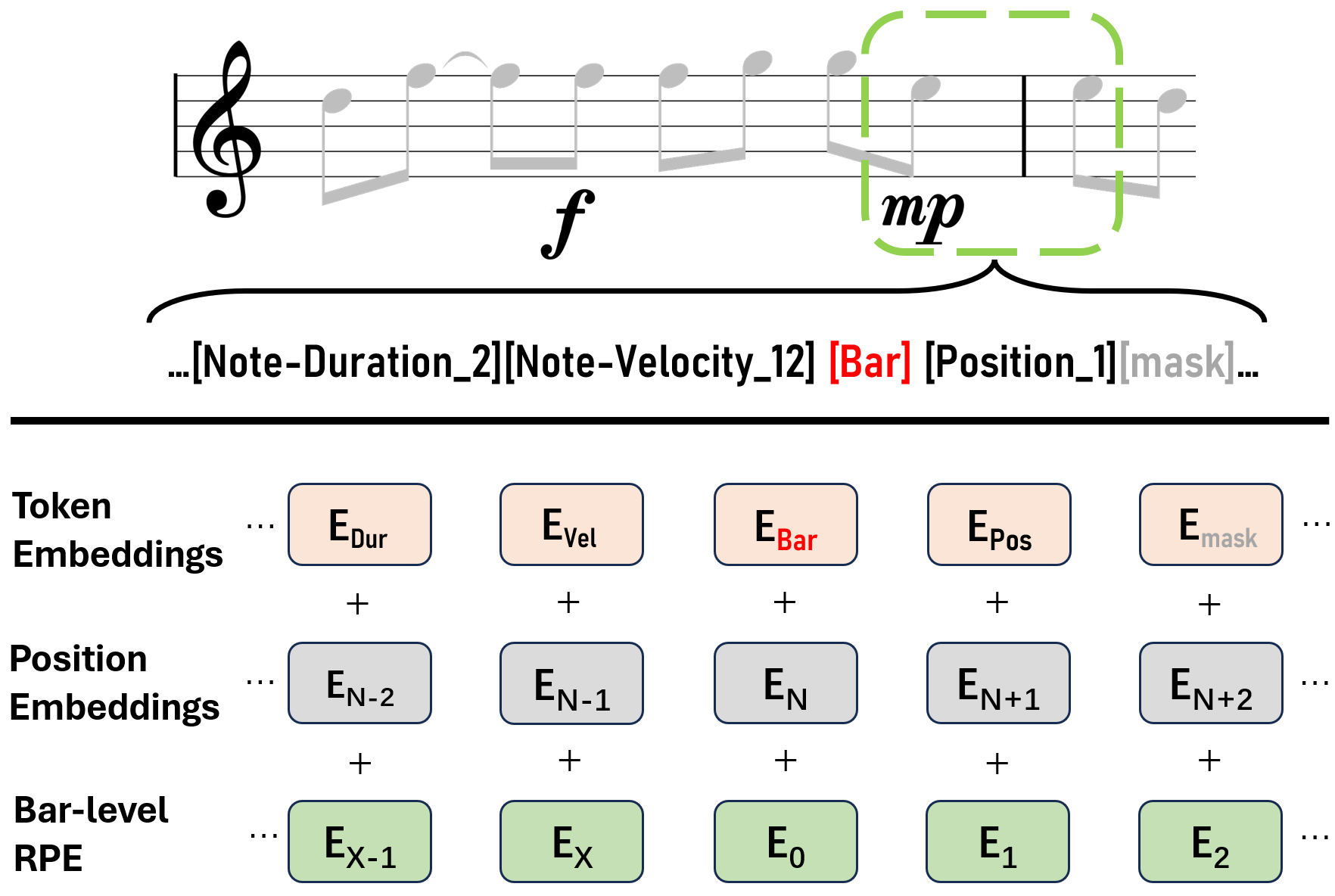}
\caption{
Illustration of the proposed bar-level relative positional encoding (RPE).
The relative position accumulates from the previous \texttt{[Bar]} token to the next \texttt{[Bar]} token,
implemented by learnable embedding, and then added to the token embedding with the vanilla positional embedding.} \label{relative position}
\end{figure}

In addition to melody, rhythm is another crucial property of music, as it reflects the progression of notes and variations in velocity, governing the dynamics of music~\cite{rhythmtheory2}.
To improve the quality in terms of rhythm, we design a fine-grained rhythm discriminator.

To facilitate the discriminator to focus on rhythms instead of other music elements, 
we decouple rhythm information from the music by 
replacing the 
\texttt{[Note-On-Pitch]} token with the \texttt{[Mask]} token.
Apart from that, we observe that the symbol ``bar'' plays a fundamental role in organizing and structuring music, which therefore can help establish the rhythmic framework of the music~\cite{rhythmtheory}. 
Based on this observation, we introduce a bar-level relative positional encoding as shown in Fig.~\ref{relative position}.
It accumulates position starting from the beginning of each bar and resets at the beginning of the next bar, \ie, from \texttt{[Bar]} token to next \texttt{[Bar]} token, embeds the bar-level relative position information into the decoupled music rhythm sequence.
%
Like other relative position embedding implementations, our bar-level position embedding is also learnable. 
%
The general position encoding of a symbolic music token, e.g., $t$-th in the whole sequence and $x$-th within the current bar, is defined as follows:

\vspace{-0.3cm}
\begin{small}
\begin{equation} 
\label{rhythm discriminator adversarial loss}
\setlength{\abovedisplayskip}{6pt}
\setlength{\belowdisplayskip}{6pt}
   A_{t,x} =cos/sin(t/1000^{2i/d})+W_{BRPE}[\delta(x, 1), \delta(x, 2), \dots],
\end{equation}
\end{small}

\noindent where the first part is traditional sine and cosine position encoding in the Transformer, the $W_{BRPE}$ is a learnable matrix, and $\delta(\cdot)$ is dirichlet function.
The rhythm discriminator shares similar back-propagation and adversarial loss to the generator as the melody discriminator in Equation (\ref{pitch discriminator adversarial loss}).


\section{EXPERIMENTS}
\label{sec:pagestyle}
\subsection{Experimental Setting}
\label{3.1 Experimental Setting}


\textbf{Dataset and preprocess.} We employ the POP909 dataset~\cite{pop909} for performance evaluation.
There are three separate tracks in each arrangement in the dataset: MELODY, BRIDGE and PIANO.
To encode a MIDI file into a sequence of discrete tokens, we adopt the REMI-like~\cite{pop_music} encoding method.
In detail, we use metric-related tokens \texttt{[Bar]}, \texttt{[Tempo]}, \texttt{[Position]} and note-related tokens \texttt{[Note-On-Pitch]}, \texttt{[Note-Duration]} and \texttt{[Note-Velocity]} to represent music, as shown in the generator part of Fig.~\ref{method}.
For fair comparisons, we retrain all the baseline models using the same data as ours, and reserve 4\% of them only for evaluation where all models take the same music piece as the condition or the prefix sequence.


\vspace{0.1cm}
\noindent \textbf{Implementation Details.}   The proposed melody and rhythm discriminators use a  6-layer encoder-only Transformer as the backbone.
Both of them have 8 heads for multi-head attention, 256 hidden dimensions, 1,024-dim feed-forward layers, and ReLU as the activation function.
In the first stage, we pre-train the generator along with all baseline models using Adam optimizer ($\beta_1$=0.9 and $\beta_2$=0.99)~\cite{adam} until the training NLL loss model below 0.55. 
Afterward, we pre-train the melody and rhythm discriminator using the dataset and the output of the trained generator for 120 epochs.
During adversarial training, both $\alpha$ and $\beta$ are set to 0.15, and using the same optimizer in the first stage to train the generator for 100 epochs.


\vspace{0.1cm}
\noindent \textbf{Baselines.}    \textbf{1) GT}~\cite{pop909}: the above-mentioned 4\% of the dataset which is not included in the training set or validation set. 
\textbf{2) Music Transformer (MT)}~\cite{music_transformer}: pioneer algorithm that successfully applied the transformer model to the domain of symbolic music generation.
\textbf{3) Theme Transformer (TT)}~\cite{theme}: a theme-conditioned music generation model optimized by NLL loss only.
\textbf{4) Anticipatory Music Transformer (AMT)~\cite{ATransformer}}: the current state-of-art model for piano music generation based on transformer. 
\textbf{5) WGAN}~\cite{adtransformer}: music generation model that utilizes a conventional
 global discriminator which will primarily serve to validate the effectiveness of our proposed fine-grained discriminator approach.
\textbf{6,7) Ours (wRo) \& Ours (wMo)}: our model uses only rhythm discriminator or melody discriminator in the adversarial training.
\textbf{8) Ours}: the complete fine-grained discriminator model.

\subsection{Objective Evaluation}
\label{3.1 Objective Evaluation}
\noindent \textbf{Evaluation Metrics.} 
We employ various metrics to demonstrate the comprehensive performance of the models.
First, following ~\cite{jazz,multitrack}, we adopt \textbf{1) pitch class entropy, 2) scale consistency}, and \textbf{3) groove consistency} to evaluate entropy of the normalized note pitch class histogram, largest pitch-in-scale rate over all major and minor scales, and mean hamming distance of the neighboring measures.
Then, we calculate the \textbf{4) pitch and 5) velocity divergence} between the generated and real music to measure the distribution similarity in melody and rhythm respectively.
Furthermore, we utilize a pre-trained music understanding model MIDI-BERT~\cite{chou2021midibert} and transform the music into feature vectors. 
The cosine \textbf{6) MIDI-BERT similarity} between generated and real music can measure the proximity of generated music to real music in a higher-level feature.

\begin{table}[htbp] 
    \caption{The results of objective evaluation. For the pitch class entropy,  groove consistency and scale consistency, a closer value to that of ground truth is considered better. Mean values
and 95\% confidence intervals are reported. Red and blue fonts denote the best and second-best performance, respectively. }
    \scalebox{0.8}{
    \begin{tabular}{l c c c c c c}
        \toprule
         & \makecell{Pitch Class \\ Entropy} & \makecell{Groove \\ Consistency} & \makecell{Scale \\ Consistency} & \makecell{Pitch \\ Divergence$\downarrow$} & \makecell{Velocity \\ Divergence$\downarrow$} & \makecell{MIDI-BERT \\ Similarity$\uparrow$}
        \\
        \midrule
        GT~\cite{pop909} & \small{$2.7726\pm0.0012$} & \small{$0.9889\pm0.0023$} & \small{$0.9799\pm0.0029$} & - & - & -\\
        \midrule
        MT~\cite{music_transformer} & \small{$2.5907\pm0.0035$} & \color{red}{\textbf{\small\bm{{$0.9876\pm0.0029$}}}} & \small{$0.9634\pm0.0039$} & \small{$0.7092\pm0.0085$} & \small{$0.3529\pm0.0089$} & \small{$0.3073\pm0.0046$}\\
        TT~\cite{theme} & \small{$2.6749\pm0.0073$} & \small{$0.9572\pm0.0038$} & \small{$0.9706\pm0.0020$} & \small{$0.1470\pm0.0007$} & \small{$0.0904\pm0.0011$} & \small{$0.2809\pm0.0027$}\\
        AMT~\cite{ATransformer} & \small{$2.7133\pm0.0094$}& \small{$0.9165\pm0.0043$} & \small{$0.9792\pm0.0033$} & \small{$1.3645\pm0.0165$} & \small{$0.6346\pm0.0132$} & \small{$0.2921\pm0.0025$}\\
        WGAN~\cite{transformer-gan} & \small{$2.6437\pm0.0129$} & \small{$0.9575\pm0.0064$} & \small{$0.9739\pm0.0074$} & \small{$0.1516\pm0.0012$} & \small{$0.0824\pm0.0010$} & \small{$0.2733\pm0.0012$}\\
        Ours (wRo) & \small{$2.7123\pm0.0088$} & \small{$0.9579\pm0.0020$} & \small{$0.9735\pm0.0047$} & \small{$0.1598\pm0.0028$} & \color{red}{\small{\bm{$0.0625\pm0.0012$}}} & \small{$0.3103\pm0.0030$}\\
        Ours (wMo) & \color{red}{\small{\bm{$2.7590\pm0.0021$}}} & \small{$0.9553\pm0.0045$} & \color{blue}{\small{\bm{$0.9743\pm0.0035$}}} & \color{blue}{\small{\bm{$0.1368\pm0.0014$}}} & \small{$0.0726\pm0.0031$} & \color{blue}{\small{\bm{$0.3205\pm0.0028$}}}\\
        Ours & \color{blue}{\small{\bm{$2.7164\pm0.0024$}}} & \color{blue}{\small{\bm{$0.9583\pm0.0022$}}} & \color{red}{\small{\bm{$0.9763\pm0.0026$}}} & \color{red}{\small{\bm{$0.1282\pm0.0013$}}} & \color{blue}{\small{\bm{$0.0675\pm0.0021$}}} & \color{red}{\small{\bm{$0.3239\pm0.0026$}}}
        \\
        \toprule
    \end{tabular}}
    \label{objective eval}
\end{table}

\noindent \textbf{Compared with SOTA methods.} 
Table \ref{objective eval} shows the comparison results. We can observe that the introduction of the fine-grained melody discriminator makes our model closer to the ground truth on pitch class entropy, scale consistency, and pitch distribution divergence.
Regarding rhythm, we can see that our model and Music Transformer~\cite{music_transformer} outperform other models in groove consistency, suggesting better rhythm control and more stable groove.
With the inclusion of the rhythm discriminator, music generated by our model is also closer to ground truth in note velocity distribution.
In terms of MIDI-BERT similarity, which largely reflects the overall quality of the generated music, our complete model achieves the highest average similarity. 
This suggests that according to the pre-trained music understanding model, music generated by our model exhibits a closer resemblance to human music compositions, both in style and musical quality.

Overall, compared to the conventional GAN-based baseline model WAGN~\cite{transformer-gan} and other benchmark models, our model achieves superior performance in objective performance metrics attributable to the fine-grained tuning of the generator by melody and rhythm discriminators.

\subsection{Subjective Evaluation}
\label{3.2 Subjective Evalution}
To assess the quality of music samples generated by our model, we conduct a listening test with 17 survey participants.
Ten of them can play at least one musical instrument and understand basic music theory.
We provide 6 sets of 30 music samples for participants, consisting of ground truth and samples generated by each model. 
All generated MIDI files are rendered to audio using MuseScore General SoundFont~\cite{multitrack}.
In the questionnaire, each participant is asked to listen to all 30 samples and then rate them on a scale of 1 to 5 according to three criteria—\emph{coherence}, \emph{richness}, and \emph{overall}. Results are reported in Table~\ref{subjective eval}.

\begin{table}[t] 
    \caption{The results of the subjective evaluation. Mean values and 95\% confidence intervals are reported.}
    \centering

    \begin{tabular*}{0.8\linewidth}{@{\extracolsep{\fill}} l c c c }
        \toprule
         & Coherence & Richness & Overall\\
        \midrule
        GT~\cite{pop909} & $3.98\pm0.18$ & $4.15\pm0.23$ & $4.06\pm0.10$ \\
        \midrule
        MT~\cite{music_transformer} & $3.56\pm0.24$ & $2.94\pm0.18$ & $3.37\pm0.33$ \\
        TT~\cite{theme} & $3.21\pm0.20$ & $3.48\pm0.16$ & {$3.44\pm0.32$} \\
          AMT~\cite{ATransformer} & $3.43\pm0.34$ & $3.32\pm 0.27$ & $3.39\pm0.24$\\
         WGAN~\cite{transformer-gan} & $3.60\pm0.35$ & $3.57\pm 0.29$ & $3.59\pm0.26$\\

        Ours & \bm{$3.68\pm0.28$} & \bm{$3.81\pm0.25$} & \bm{$3.71\pm0.26$}\\
        \toprule
    \end{tabular*}
    \label{subjective eval}
\end{table}

The results show that our model achieves higher scores across all criteria than other models.
It's worth noticing that while Music Transformer~\cite{music_transformer} surpasses our model in terms of the groove consistency metric in objective evaluation, it is less favorable than our model in terms of coherence and richness in subjective listening tests, especially richness.
Based on the feedback from survey participants, we find that the music generated by Music Transformer contains a larger amount of repetition, leading to a monotonous listening experience.
%
Benefiting from the fine-grained adversarial optimization, our model outperforms the SOTA single discriminator method WGAN.
The performance improvements demonstrate the effectiveness of our method on both coherence and richness aspects and overall quality.

%



\subsection{Ablation Studies and Qualitative Evaluation}
\label{ssec:subhead}

As shown in Table~\ref{objective eval}, when using only the fine-grained melody discriminator, our method has shown a significant improvement compared to other baseline models in metrics strongly related to melody such as pitch class entropy, scale consistency, and pitch divergence, reaching a closer level to real music. 
Moreover, since melody and pitch are the core components of music expression~\cite{ecvae}, the melody discriminator enables the model to generate more realistic music, as indicated by the outstanding MIDI-BERT similarity.
When solely using the fine-grained rhythm discriminator, our method also achieves better performance than baseline models in velocity divergence and MIDI-BERT similarity, proving the effectiveness of both fine-grained discriminators.

\begin{figure*}
	\centering  
	\subfigbottomskip=1pt 
	\subfigcapskip=-3pt 
	\subfigure[Note Pitch Distribution]{
		\includegraphics[width=0.48\linewidth]{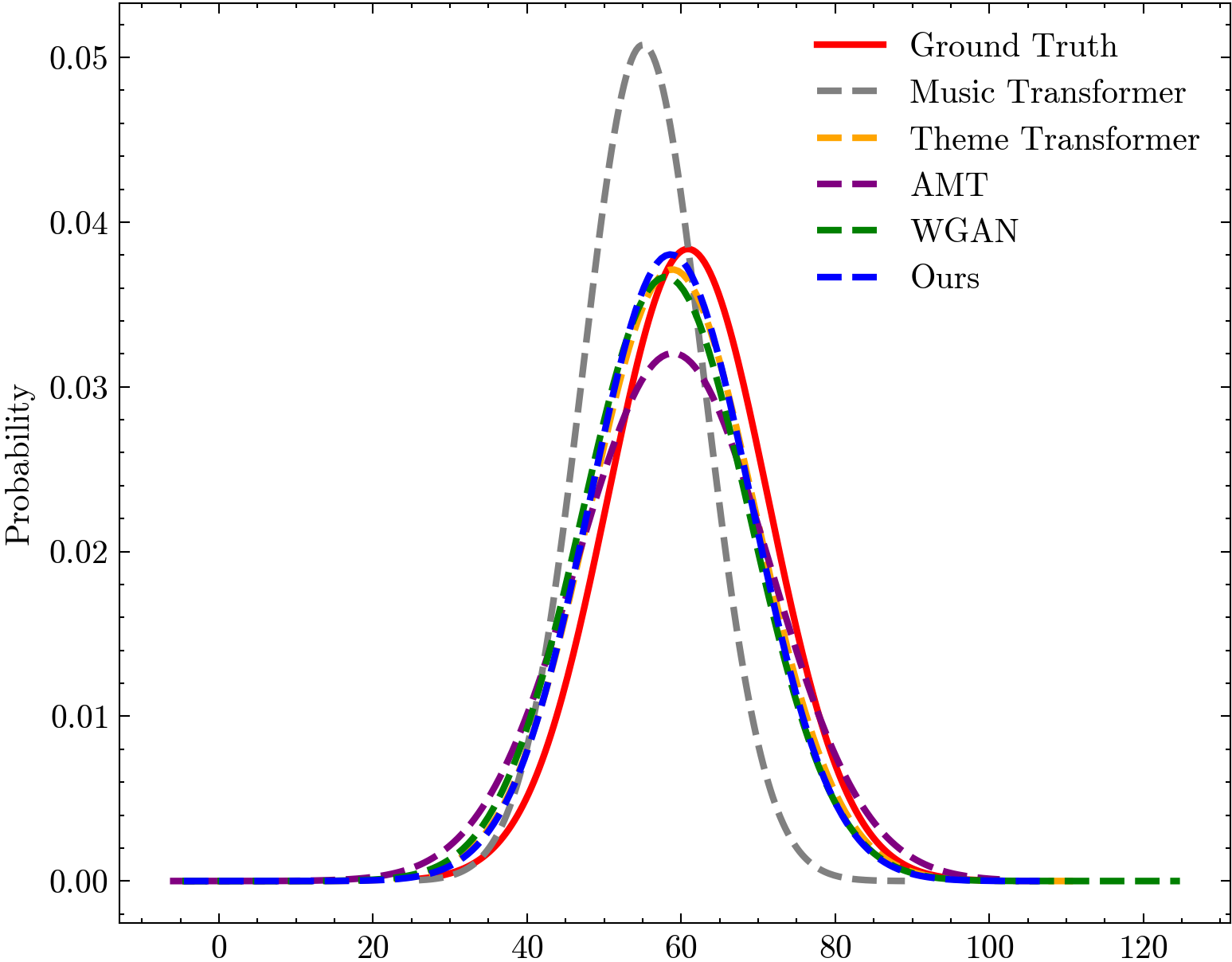}}
	\subfigure[Note Velocity Distribution]{
		\includegraphics[width=0.48\linewidth]{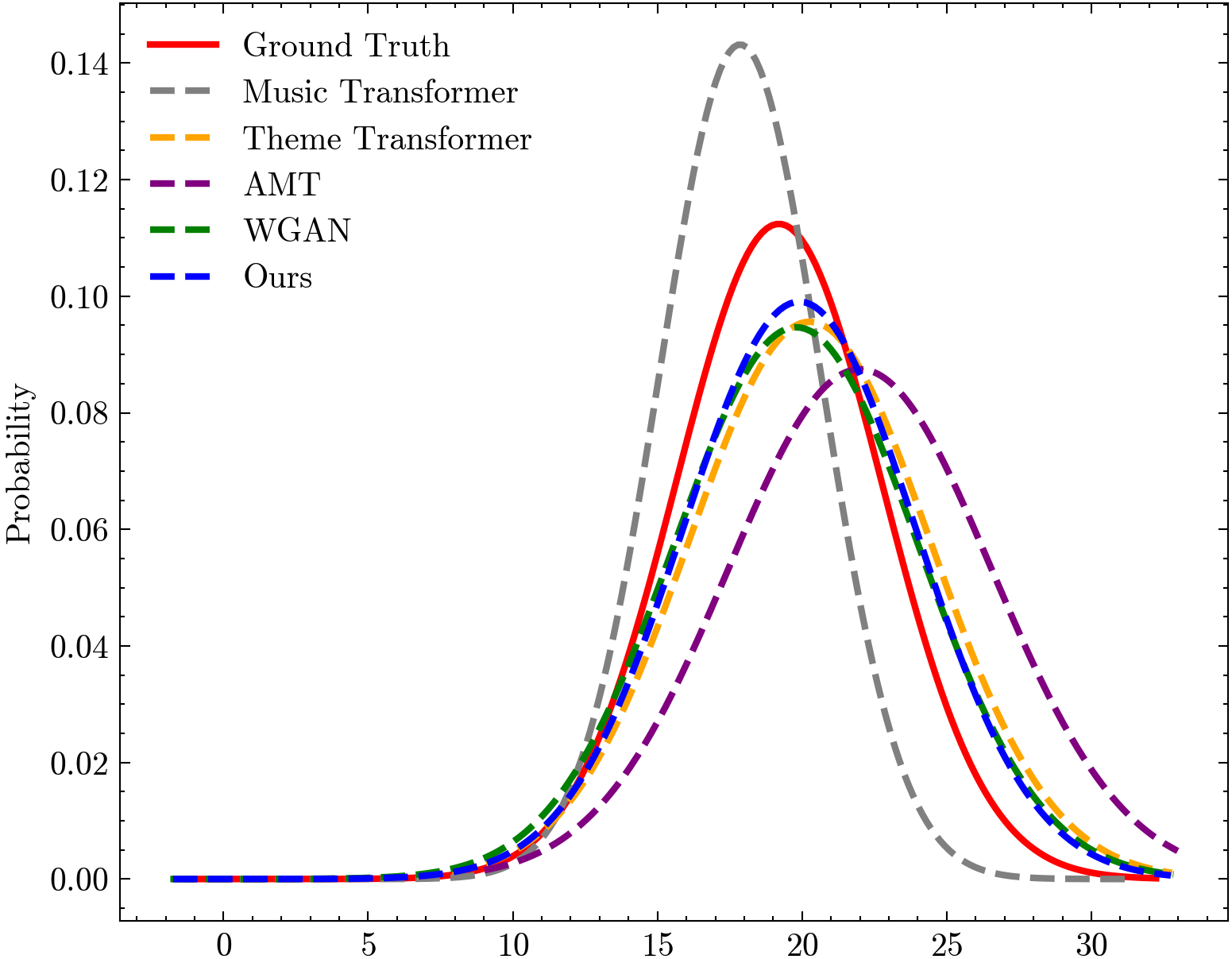}}
	\caption{
                Quantitative analysis. 
                (a) \& (b)Visualization of the note pitch and note velocity distribution of music generated by different models and the Ground Truth. 
    }
    \label{visual_distribution}
\end{figure*}

Fig.~\ref{visual_distribution} (a) and (b) visualize the pitch and velocity distribution between the music generated by different models and the ground truth.
We approximate each distribution to a normal distribution for better comparison. 
It can be observed that benefiting from the fine-grained melody and rhythm discriminators, our model is closer to real music in both attributes.

\begin{figure}
\centering
\includegraphics[width=0.6\linewidth]{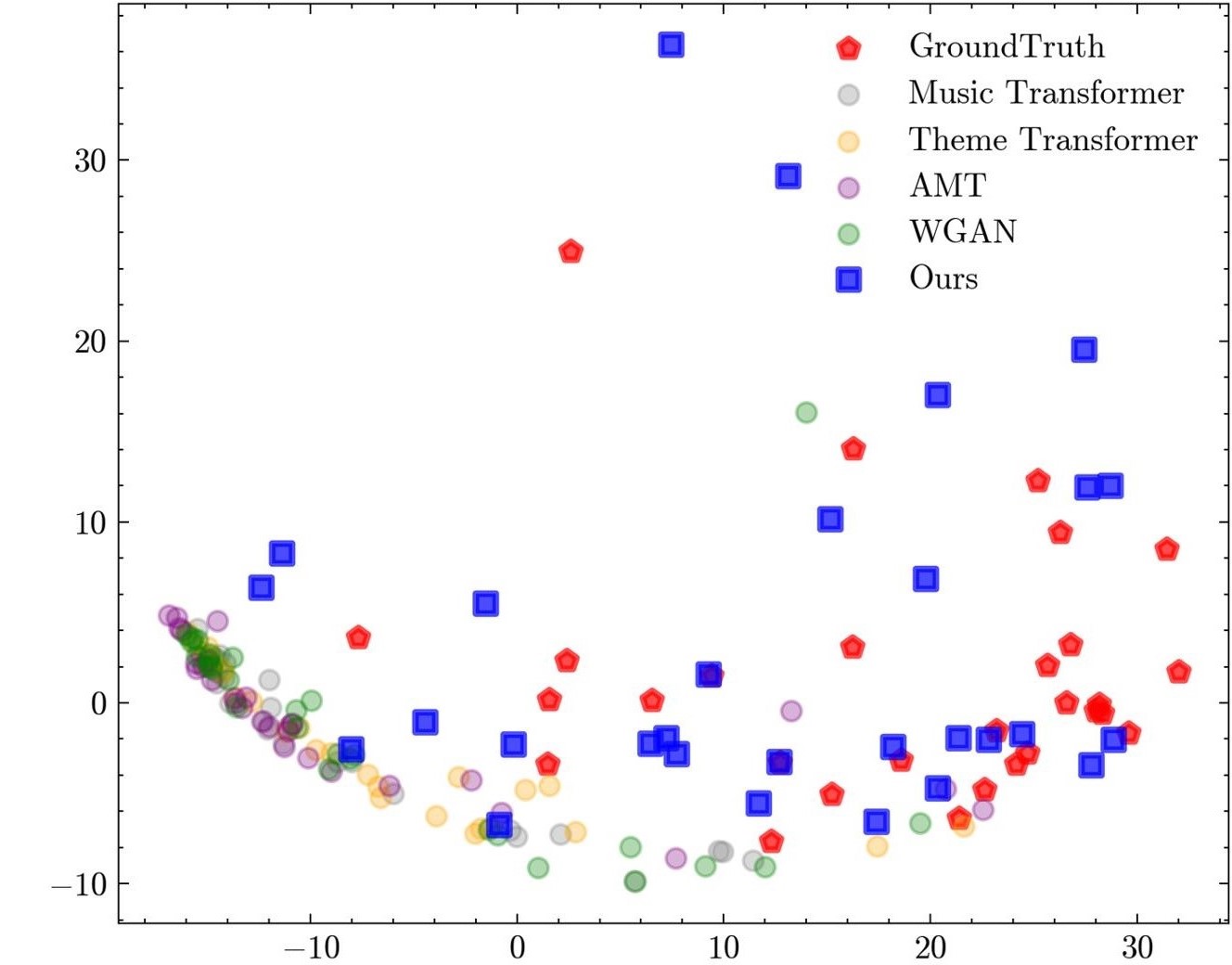}
\caption{The PCA visualization results of music feature obtained from MIDI-BERT~\cite{chou2021midibert}.
} 
\vspace{-0.4cm}
\label{distribution}

\end{figure}

To better evaluate the effectiveness of our method, we compute the feature vectors of music generated by each model along with the ground truth using the pre-trained music understanding model MIDI-BERT~\cite{chou2021midibert}.
We utilize PCA algorithm to reduce the high-dimensional feature vectors to 2 dimensions and visualize the feature vectors in Fig.~\ref{distribution}.
It is evident that the music generated by other baseline models exhibits a noticeable domain gap compared to the ground truth, while MIDI-BERT can effectively distinguish whether they are real music or not. 
Music Transformer~\cite{music_transformer} and AMT~\cite{ATransformer} are trained solely using maximum likelihood as a training objective function, thus they suffer from quality degradation caused by exposure bias.
Therefore, music generated by them has a certain degree of uniformity, with their feature vector distributions being highly concentrated.
\begin{figure}
\centering
\includegraphics[width=0.8\linewidth]{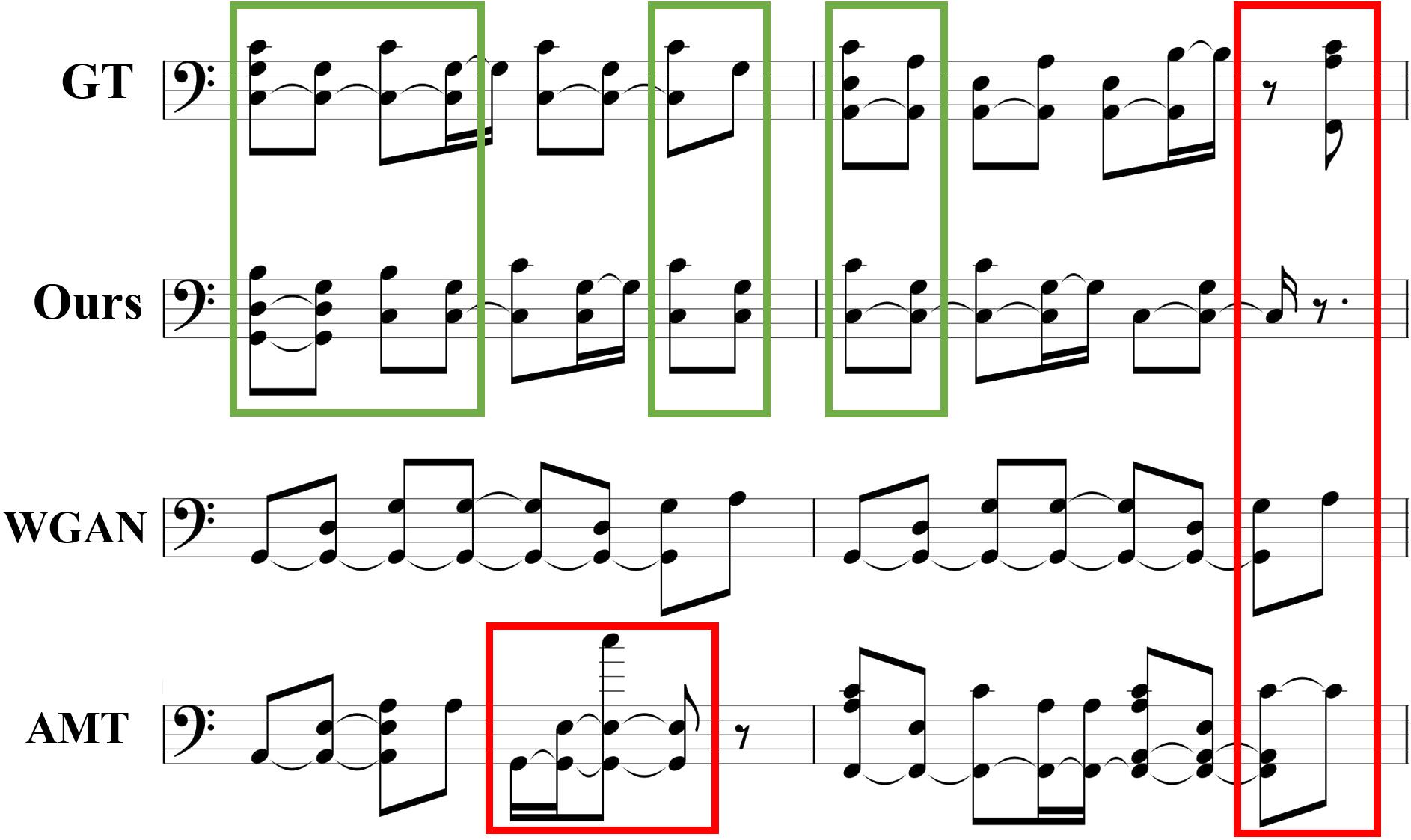}
\caption{WGAN~\cite{transformer-gan}, AMT~\cite{ATransformer}, and our model’s generated examples and their corresponding ground truth music piece.
} 
\label{exp_score}
\vspace{-0.4cm}
\end{figure}
WGAN~\cite{transformer-gan} alleviates quality degradation through adversarial loss, resulting in a more dispersed distribution of its feature vectors compared to the previous two, but still exists a considerable gap from the ground truth.
Equipped with fine-grained discriminators, our model's style vectors exhibit a distribution that closely resembles the ground truth and also diversifies itself.


Fig.~\ref{exp_score} illustrates examples generated by both our model, WGAN~\cite{transformer-gan} and AMT~\cite{ATransformer} when giving the same condition.
Compared to other models, our generated example exhibits a closer resemblance to the ground truth in terms of melody design and transitions (highlighted by the green bounding boxes).  
The example from AMT~\cite{ATransformer} and WGAN~\cite{transformer-gan} exhibits some disharmony caused by abnormal notes and discrepant rhythm patterns (highlighted by the red bounding box).

\section{Conclusion}
\label{sec:typestyle}
This work proposes a fine-grained discriminators architecture for the symbolic music generation task.
We decouple the music into melody and rhythm for independent discrimination, which provides the generator with more specific feedback.
We also devise a pitch augment strategy and a bar-level relative positional encoding scheme to enhance the learning of melody discriminator and rhythm discriminator, respectively. 
Extensive objective and subjective results demonstrate the effectiveness of the proposed method. 

\bibliographystyle{splncs04}
\bibliography{ref}
%
%
%
%




\end{document}